\documentclass[aps,pra,twocolumn,showpacs,superscriptaddress,longbibliography]{revtex4-2}
\usepackage{graphicx} 
\usepackage{amsmath}
\usepackage{graphicx,epstopdf}
\usepackage{gensymb}
\newcommand{\be}{\begin{equation}}
\newcommand{\ee}{\end{equation}}
\newcommand{\bea}{\begin{eqnarray}}
\newcommand{\eea}{\end{eqnarray}}
\newcommand{\bse}{\begin{subequations}}
	\newcommand{\ese}{\end{subequations}}

\usepackage{color}
\usepackage[colorlinks,bookmarks=false,citecolor=darkblue,linkcolor=red,urlcolor=blue]{hyperref}

\definecolor{darkred}{rgb}{0.7,0.0,0.0}

\definecolor{darkblue}{rgb}{0,0.02,0.45}

\definecolor{darkgreen}{rgb}{0.02,0.45,0.0}

\definecolor{violet}{rgb}{0.8,0.2,0.6}

\begin{document}
\title{Double magnetic transitions and exotic field induced phase in the triangular lattice antiferromagnets Sr$_3$Co(Nb,Ta)$_2$O$_9$}

\author{Surender Lal}
\author{Sebin J Sebastian}
\author{S. S. Islam}
\affiliation{School of Physics, Indian Institute of Science
	Education and Research Thiruvananthapuram-695551, India}
\author{M. P. Saravanan}
\affiliation{Low Temperature Laboratory, UGC-DAE Consortium for Scientific Research, University Campus, Khandwa Road, Indore 452001, India}
\author{M. Uhlarz}
\author{Y. Skourski}
\affiliation{Dresden High Magnetic Field Laboratory (HLD-EMFL), Helmholtz-Zentrum Dresden-Rossendorf, 01328 Dresden, Germany}
\author{R. Nath}
\email{rnath@iisertvm.ac.in}
\affiliation{School of Physics, Indian Institute of Science Education and Research Thiruvananthapuram-695551, India}
\date{\today}
	
\begin{abstract}
Two triangular lattice antiferromagnets Sr$_3$Co(Nb,Ta)$_2$O$_9$ with an effective $j_{\rm eff}=1/2$ of Co$^{2+}$ are synthesized and their magnetic properties are investigated via magnetization and heat capacity measurements. The leading in-plane antiferromagnetic exchange coupling is estimated to be $J/k_{\rm B} \simeq 4.7$~K and 5.8~K, respectively. Both the compounds feature two-step magnetic transitions at low temperatures [($T_{\rm N1} \simeq 1.47$~K and $T_{\rm N2} \simeq 1.22$~K) and ($T_{\rm N1} \simeq 0.88$~K and $T_{\rm N2} \simeq 0.67$~K), respectively], driven by weak easy-axis anisotropy. Under magnetic field Sr$_3$CoNb$_2$O$_9$ evinces a 
plateau at $1/3$ magnetization. Interestingly, the high field magnetization of Sr$_3$CoTa$_2$O$_9$ reveals an exotic regime (between $H_{\rm S1}$ and $H_{\rm S2}$), below the fully polarized state in which the heat capacity at low temperatures is governed by a power law ($C_{\rm p} \propto T^{\alpha}$) with a reduced exponent $\alpha \simeq 2$. These results demonstrate an unusual field induced state with gapless excitations in the strongly frustrated magnet Sr$_3$CoTa$_2$O$_9$. The complete $T-H$ phase diagram is discussed for both the compounds.
\end{abstract}

\maketitle

\section{Introduction}
Geometrically frustrated magnets have attracted a revived interest since frustration effect may cause the absence of magnetic long-range-order (LRO), leading to an abundance of novel states of matter~\cite{Lacroix2011,Diep2005}. The size of local magnetic moment is also a convenient tuning parameter that controls the magnitude of quantum fluctuations and has broad implications on the ground state properties. For instance, a reduced spin value, especially spin-$1/2$ amplifies the effect of quantum fluctuations and precipitates more non-trivial ground states. A renowned testimony of magnetic frustration and quantum fluctuations is the quantum spin-liquid (QSL), a highly entangled and dynamically disordered many-body state~\cite{Balents199,*Savary016502}. 
Over the years, relentless efforts are made to experimentally devise appropriate model compounds with spin-$1/2$ that may promote this disordered state.

Frustrated spin-$1/2$ triangular lattice antiferromagnet (TLAF) is widely believed to be a model system to host QSL driven by ground state degeneracy in two-dimension (2D)~\cite{Anderson153}. In an isotropic Heisenberg TLAF with only nearest-neighbour (NN) interaction, the spins order antiferromagnetically forming 120$^{\circ}$ spin structure in zero magnetic field, known as 3-sublattice N$\acute{e}$el state~\cite{Singh1766,Capriotti3899,White127004}. When external magnetic field is applied, the N$\acute{e}$el order is subverted and an "up-up-down" ($uud$) configuration is stabilized over a wide field range before reaching saturation. This results in a magnetization plateau at $1/3$ of the magnetic saturation stemming from quantum and/or thermal fluctuations~\cite{Chubukov69,Kamiya2666}. Magnetic anisotropy and interactions beyond nearest-neighbour are also two crucial parameters, inherently present in majority of the experimental systems, influence the ground state significantly and give rise to more complex low temperature phases, including QSL~\cite{Maksimov021017,Hu140403,Melchy064411,Ranjith094426,*Ranjith024422}.

The Co$^{2+}$ ($3d^7$)-based TLAFs are a special class of compounds that manifests various exotic phases of matter similar to the $4f$ systems (e.g. Ce$^{3+}$ and Yb$^{3+}$)~\cite{Li167203,Bordelon094421,Somesh2023}. In most of these compounds, the impact of crystal electric field (CEF) and spin-orbit coupling (SOC) in a non-cubic environment lead to a Kramers doublet and the effective
magnetic moment of Co$^{2+}$ ions, which possess the true
spin $S=3/2$, can be described by the pseudo-spin $j_{\rm eff}=1/2$ at
low temperatures, well below the energy scale of the spin-orbit coupling constant ($\lambda/k_{\rm B}$)~\cite{Susuki267201,Shirata057205}. This allows one to study the combined effects of magnetic frustration and quantum fluctuations due to reduced spin, at very low temperatures~\cite{Zhong14505}. In the past few years, a series of triple perovskites with general formula $A_3$Co$B_2$O$_9$ ($A$= Sr, Ba and $B=$~Sb, Ta, Nb) have been rigorously investigated in which the magnetic Co$^{2+}$ ions are embedded onto 2D triangular layers, separated by layers of non-magnetic $A$ and $B$ atoms~\cite{Shirata057205,Susuki267201,Lee104420,Yokota014403,Ranjith115804}.
The most celebrated compound in this family is Ba$_3$CoSb$_2$O$_9$ which shows successive magnetic transitions, magnetization plateau at $1/3$ and $3/5$ of saturation magnetization, and very recently the QSL is claimed~\cite{Zhou267206,Susuki267201,Kamiya2666}.
Several other Co$^{2+}$ based $j_{\rm eff} =1/2$ TLAFs are also reported to show diverse physics with complex magnetic orderings~\cite{Li4216,Kojima174406,Rawl174438,Zhong224430,Rawl060412,Muthuselvam174430}. Because of the localized nature of the $3d$ electrons, cobaltates featuring honeycomb lattice also offer a promising ground to look for Kitaev spin-liquid ~\cite{Liu014407,Sano014408,Liu047201}.
Indeed, a field induced Kitaev spin-liquid like behaviour has been observed in Na$_2$Co$_2$TeO$_6$~\cite{Lin5559} and BaCo$_{2}$(AsO$_{4}$)$_{2}$~\cite{Zhong6953}. However, their complete phase diagram remains obscure.

\begin{figure}
	\centering
	\includegraphics[width= \linewidth] {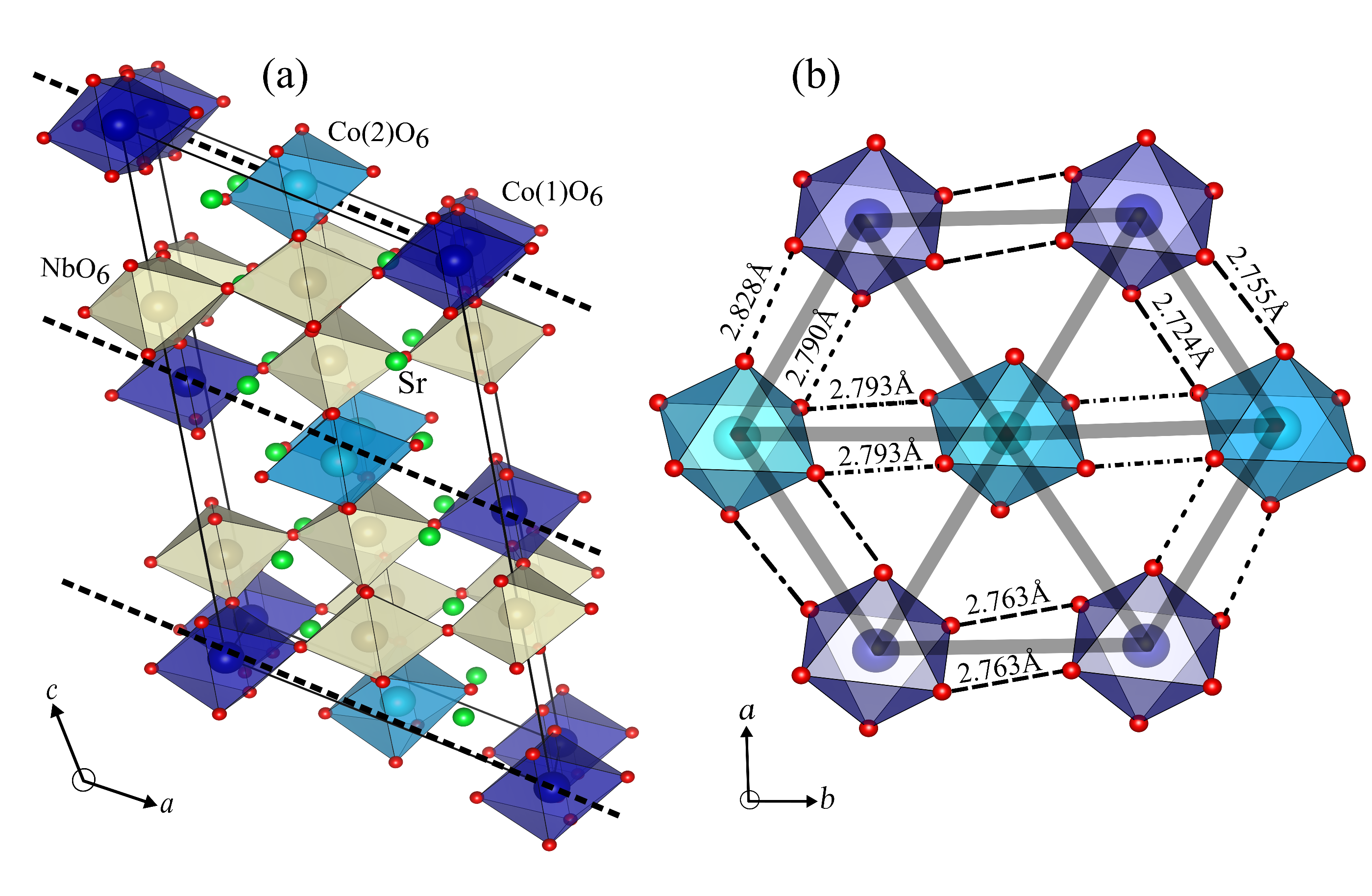}
	\caption{(a) A three-dimensional view of the crystal structure of Sr$_3$CoNb$_2$O$_9$. The dashed lines guide the triangular layers composed of corner shared CoO$_6$ and NbO$_{6}$ octahedra. The Co(1)O$_6$ and Co(2)O$_6$ octahedra are shown in different colours. (b) A section of the triangular layer that depicts nearly parallel edge-shared arrangement of CoO$_6$ octahedra and presents the honeycomb lattice, after removing NbO$_{6}$ octahedra.}
	\label{Fig1}
\end{figure}
In this work, we report a comprehensive study of the synthesis and thermodynamic properties of two new frustrated $j_{\rm eff} = 1/2$ iso-structural TLAFs Sr$_3$CoNb$_2$O$_9$ and Sr$_3$CoTa$_2$O$_9$. Sr$_3$CoNb$_2$O$_9$ is reported to crystallize in a monoclinic structure with space group $P2_{1}/c$~\cite{Lee476004,Ting2295}. Its crystal structure is illustrated in Fig.~\ref{Fig1}.
There are two inequivalent Co atoms residing at the Co(1) $2a$(0,0,0) and Co(2) $2d$(1/2,1/2,0) sites, respectively which are coordinated with O atoms forming slightly distorted CoO$_{6}$ octahedra. Two-dimensional (2D) triangular layers are formed by the corner sharing of magnetic CoO$_{6}$ and non-magnetic NbO$_6$ octahedra in the $ab$-plane. Figure~\ref{Fig1}(b) presents nearly parallel edge-sharing CoO$_6$ octahedra (NbO$_6$ omitted) displaying possible superimposed honeycomb lattices. The non-magnetic Sr atoms are located at the interstitial positions. In each layer, an isosceles triangular unit is made up of either one Co(1)$^{2+}$ and two Co(2)$^{2+}$ or two Co(1)$^{2+}$ and one Co(2)$^{2+}$ ions.
Moreover, the Co$^{2+}$ triangular layers are arranged in an $AAA$-type stacking perpendicular to the $c$-axis which results in minuscule inter-layer frustration~\cite{Liu224413}.
Magnetic measurements reveal double transitions at low temperatures typical for compounds with easy-axis anisotropy. Despite structural similarity, Sr$_3$CoNb$_2$O$_9$ exhibits a $1/3$ magnetization plateau which is absent for Sr$_3$CoTa$_2$O$_9$. Further, Sr$_3$CoTa$_2$O$_9$ manifests an extended field induced critical regime where the ground state appears to be of QSL type while this regime is found to be narrow for Sr$_3$CoNb$_2$O$_9$.

\section{EXPERIMENTAL DETAILS}
Polycrystalline samples of Sr$_3$CoNb$_2$O$_9$ and Sr$_3$CoTa$_2$O$_9$ were synthesized by the traditional solid-state reaction method. Stoichiometric amount of SrCO$_3$ (99.99\%, Sigma Aldrich), CoO (99.999\%, Sigma Aldrich), and Nb$_2$O$_5$/Ta$_2$O$_5$ (99.999\%, Sigma Aldrich) were mixed and ground thoroughly for three hours. The mixtures were pressed into disc shaped pellets and sintered at $1100~^\circ$C for 24~hours. In the next step, the heat treatment was repeated at $1250~^\circ$C for 24~hours after regrinding and re-pelletization. The powder x-ray diffraction (XRD) was recorded using a PANalytical powder diffractometer (Cu$K _{\alpha}$ radiation, $\lambda = 1.54182$~\AA) at room temperature. Rietveld refinement of the powder XRD data for both the compounds was carried out using the FULLPROF software package~\cite{Rodriguez55} to check the phase purity of the samples and to calculate the structural parameters.

The $dc$ magnetization ($M$) measurement was performed with a superconducting quantum interference device (SQUID, MPMS-3, Quantum Design) magnetometer as a function of temperature (1.8~K$\leq T \leq$~350~K) and magnetic field (0~$\leq H \leq$~7~T). High-field magnetization $M(H)$ was measured in a pulsed magnetic field at the Dresden High Magnetic Field Laboratory~\cite{Tsirlin132407,*Skourski214420}. Heat capacity ($C_{\rm P}$) as a function of $T$ (0.1~K$\leq T \leq$~300~K) and $H$ (0~$\leq H \leq$~9~T) was measured using thermal relaxation technique in a physical property measurement system (PPMS, Evercool-II, Quantum Design). For Sr$_3$CoNb$_2$O$_9$ and Sr$_3$CoTa$_2$O$_9$, the measurements were performed down to 0.4~K and 0.1~K using $^3$He and dilution inserts, respectively in PPMS.

\section{Results}
\subsection{Powder x-ray Diffraction}
\begin{figure}
	\centering
	\includegraphics[width= \linewidth] {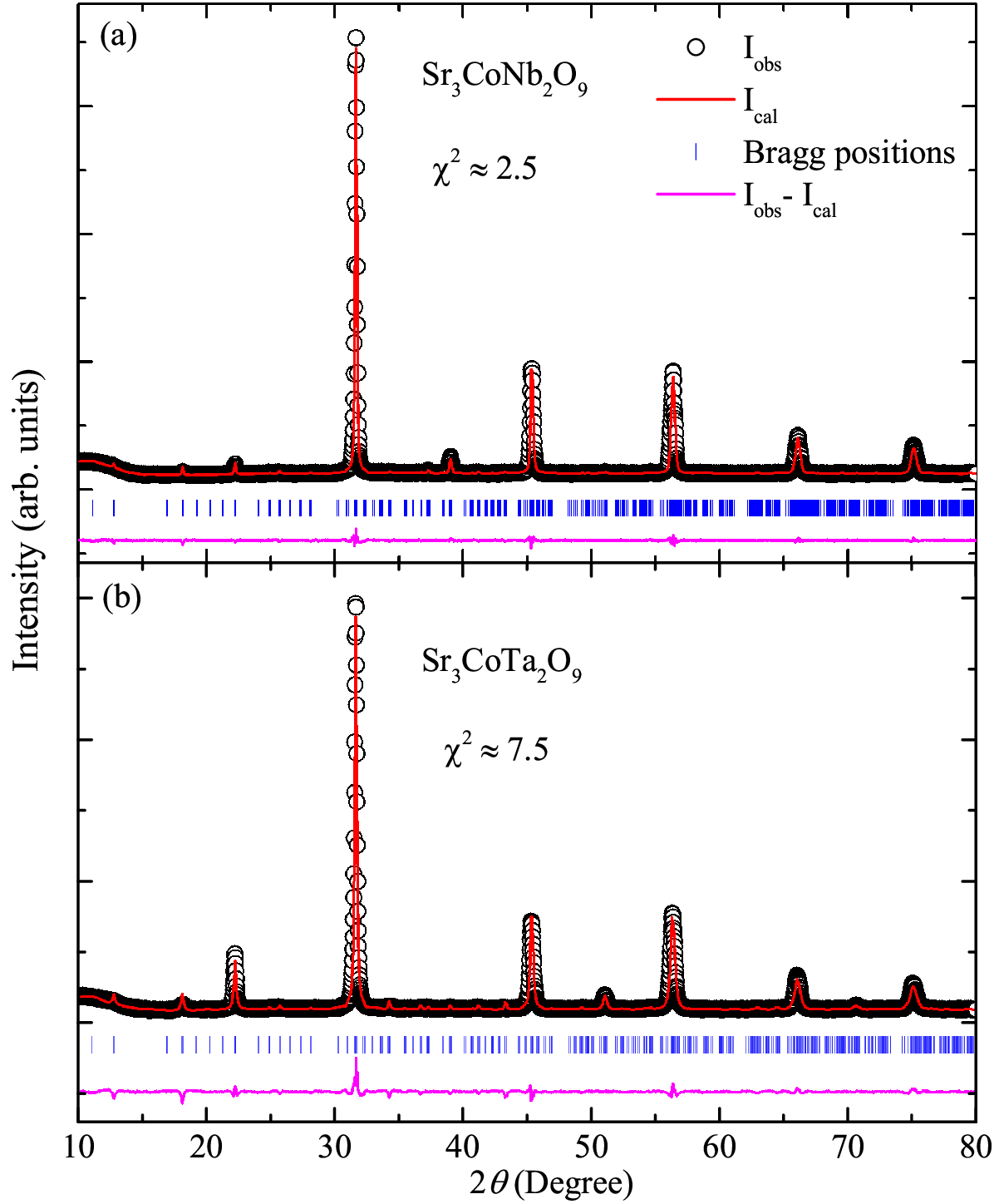}
	\caption{Room temperature powder XRD patterns of (a) Sr$_3$CoNb$_2$O$_9$ and (b) Sr$_3$CoTa$_2$O$_9$. The black circles show the XRD data and the red line represents the calculated pattern. The blue small vertical lines represent the Bragg positions. The green line at the bottom represents the difference between observed and calculated intensities.}
	\label{Fig2}
\end{figure}

\begin{table}
	\caption{Lattice parameters obtained from the Rietveld refinement of the room temperature powder XRD data of Sr$_3$Co(Nb,Ta)$_2$O$_9$ (Monoclinic, $P2_{1}/c$).}
	\label{table1}
	\begin{ruledtabular}
		\begin{tabular}{cc@{\hspace{2em}}cc@{\hspace{-2em}}cc@{\hspace{-2em}}r}
			Parameters & Sr$_3$CoNb$_2$O$_9$ & Sr$_3$CoTa$_2$O$_9$  \\
			\hline
			$a$~{(\AA)} & 9.7867(5) & 9.7790(5)  \\
			$b$~{(\AA)} & 5.6460(2) &   5.6643(4)\\
			$c$~{(\AA)} & 17.0057(4) & 16.957(1) \\
			$\beta$~{(\degree)} & 125.32(3) & 125.20(3) \\
			$V_{\rm cell}$~(\AA$^3$) & 766.67(5) & 767.46(8)  \\
			Bragg R-factor & 2.2 & 11 \\
			Rf-factor & 6.5 & 6.0 \\
			$\chi^{2}$ & 2.5 & 7.7 \\
			Co(1) - Co(1)~{(\AA)} & 5.6460(1) & 5.6644(5)  \\
			Co(2) - Co(2)~{(\AA)} & 5.6460(1) & 5.6644(5) \\
			Co(1) - Co(2)~{(\AA)} & 5.6493(3) & 5.6505(3) \\
		\end{tabular}
	\end{ruledtabular}
\end{table}

\begin{table}[]
	\caption{ Atomic positions obtained from the Rietveld refinement of the room temperature powder XRD data of Sr$_3$Co(Nb,Ta)$_2$O$_9$.}
	\label{table2}
	\begin{ruledtabular}
		\begin{tabular}{cc@{\hspace{-1em}}cc@{\hspace{1em}}cc@{\hspace{1em}}cc@{\hspace{1em}}cc@{\hspace{1em}}cc@{\hspace{2em}}r}
			Atom & Wyckoff & $x$ & $y$ & $z$ & Occ.\\
			\hline
			Sr1 & 4e & 0.2500 & 0.500 & 0.08321(6) & 1.0 \\
			&  & 0.2500 & 0.500 & 0.0803(7) & 1.0 \\
			Sr2 & 4e & 0.7500 & 0.000 & 0.08262(3) & 1.0 \\
			&  & 0.7500 & 0.000 & 0.07348(3) & 1.0 \\
			Sr3 & 4e & 0.2500 & 0.00 & 0.2500 & 1.0 \\
			&    & 0.2500 & 0.00 & 0.2500 & 1.0 \\
			Co1 & 2a & 0.000 & 0.000 & 0.000 & 0.5 \\
			&    & 0.000 & 0.000 & 0.000 & 0.5\\
			Co2 & 2d & 0.500 & 0.500 & 0.000 & 0.5 \\
			&    & 0.500 & 0.500 & 0.000 & 0.5 \\
			Nb1 & 4e & 0.50949(5) & 0.500 & 0.33735(3) & 1.0 \\
			Ta1 &    & 0.51011(3) & 0.500 & 0.33661(7) & 1.0 \\
			Nb2 & 4e & 0.00710(4) & 0.500 & 0.16336(8) & 1.0 \\
			Ta2 &    & 0.00975(4) & 0.500 & 0.16713(6) & 1.0 \\
			O1 & 4e & 0.97200(3) & 0.74250(1) & 0.2430(6) & 1.0 \\
			&    &  0.97200(3) & 0.74250(1) & 0.2430(6) & 1.0 \\
			O2 & 4e & 0.5260(3) & 0.78710(5) & 0.27160(6) & 1.0 \\
			&    & 0.5260(3) & 0.78710(5) & 0.27160(6) & 1.0 \\
			O3 & 4e & 0.2500(0) & 0.5528(4) & 0.26260(2) & 1.0 \\
			&    &0.2500(0) & 0.5528(4) & 0.26260(2) & 1.0 \\
			O4 & 4e & 0.97210(9) & 0.7330(7) & 0.90670(4) & 1.0 \\
			&    & 0.97210(9) & 0.7330(7) & 0.90670(4) & 1.0 \\
			O5 & 4e & 0.0279(6) & 0.2518(6) & 0.92020(5) & 1.0 \\
			&    & 0.0279(6) & 0.2518(6) & 0.92020(5) & 1.0 \\
			O6 & 4e & 0.4721(6) & 0.2777(0) & 0.89190(1) & 1.0 \\
			&    & 0.4721(6) & 0.2777(0) & 0.89190(1) & 1.0 \\
			O7 & 4e & 0.52790(9) & 0.79650(6) & 0.9352(5) & 1.0 \\
			&    & 0.52790(9) & 0.79650(6) & 0.9352(5) & 1.0 \\
			O8 & 4e & 0.74060(8) & 0.04280(5) & 0.9110(3) & 1.0 \\
			&    & 0.74060(8) & 0.04280(5) & 0.9110(3) & 1.0 \\
			O9 & 4e & 0.24060(7) & 0.5428(0) & 0.90780(3) & 1.0 \\
			&    & 0.24060(7) & 0.5428(0) & 0.90780(3) & 1.0 \\	
		\end{tabular}
	\end{ruledtabular}
\end{table}
Powder XRD data collected at room temperature for Sr$_3$CoNb$_2$O$_9$ and Sr$_3$CoTa$_2$O$_9$ are shown in Fig.~\ref{Fig2}(a) and (b), respectively. With the help of Rietveld refinement, all the peaks could be modeled assuming monoclinic structure ($P2_{1}/c$) for both the compounds and taking initial parameters from Refs.~\cite{Lee476004,Ting2295}. This suggests that the new compound obtained replacing Nb by Ta also stabilizes in the same crystal structure. During refinement, the positions of oxygen atoms for Sr$_3$CoTa$_2$O$_9$ couldn't be refined and kept fixed to the values of Sr$_3$CoNb$_2$O$_9$. For a comparison, the refined lattice parameters and atomic positions for both the compounds are tabulated in Table~\ref{table1} and \ref{table2}, respectively. The obtained structural parameters for Sr$_3$CoNb$_2$O$_9$ are in close agreement with the previous reports~\cite{Lee476004,Ting2295}.
 
Upon replacing Nb by Ta, the lattice constants $a$ and $c$ are found to decrease while $b$ increases. This results in an overall increase in the unit cell volume ($V_{\rm cell}$). In the crystal structure, all the magnetic Co$^{2+}$ layers are equally spaced with an inter-layer separation of $\sim 17.0057$~{\AA} and $\sim 16.9574$~{\AA} for Sr$_3$CoNb$_2$O$_9$ and Sr$_3$CoTa$_2$O$_9$, respectively.


\subsection{Magnetization}
\begin{figure}
	\centering
	\includegraphics[width= \linewidth] {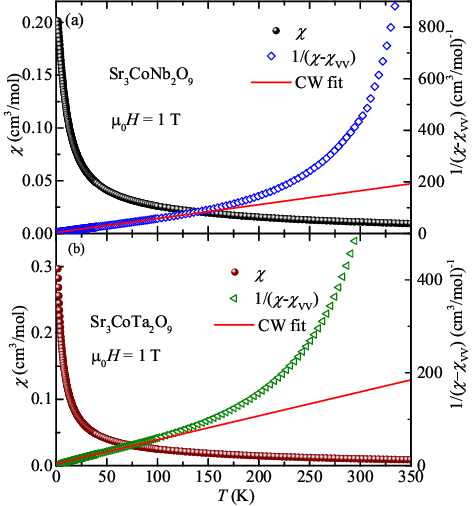}
	\caption{$\chi$ vs $T$ measured in $\mu_{0}H=1$~T for (a) Sr$_3$CoNb$_2$O$_9$ and (b) Sr$_3$CoTa$_2$O$_9$. $1/\chi$ after subtracting $\chi_{\rm VV}$ is plotted as a function of $T$ in the right $y$-axis to emphasize the low-$T$ linear portion. Solid line represents the Curie-Weiss fit in the low temperature linear regime, as discussed in the text.}
	\label{Fig3}
\end{figure} 
Temperature dependent magnetic susceptibility $\chi$ ($\equiv M/H$) of Sr$_3$CoNb$_2$O$_9$ and Sr$_3$CoTa$_2$O$_9$ measured in a magnetic field of $H=10$~kOe is shown in Fig.~\ref{Fig3}(a) and (b), respectively. As the temperature is lowered, $\chi(T)$ increases in a Curie-Weiss (CW) manner. No clear indication of any magnetic long-range ordering (LRO) is observed down to 2~K. When the inverse susceptibility ($1/\chi$) is plotted against temperature, it exhibits a linear behaviour at high temperatures and a change in slope at around $\sim 50$~K. This change in slope is a possible indication of the crossover of spin state of Co$^{2+}$ from high temperature $S=3/2$ state to an effective $j_{\rm eff} = 1/2$ ground state~\cite{Rawl060412,Li4216,Haraguchi214421,*Pietrzyk2316}. In order to extract the magnetic parameters, $1/\chi$ in the low and high temperature linear regions is fitted by the modified CW law
\begin{equation}
	\chi (T)={{\chi }_{0}}+\frac{C}{T-{{\theta}_{\rm CW}}},
	\label{CW} 
\end{equation}
where $\chi_{0}$ represents the temperature independent susceptibility, $C$ is the CW constant, and $\theta_{\rm CW}$ is the characteristic CW temperature. The fit in high temperature ($T>200$~K) range yields ($\chi_{0}\simeq 5.315 \times 10^{-4}$~cm$^3$/mol, $C \simeq  3.37$~cm$^{3}$K/mol, and $\theta_{\rm CW} \simeq -24$~K) and ($\chi_{0}\simeq 2.84 \times 10^{-4}$~cm$^3$/mol, $C\simeq 3.24$~cm$^{3}$K/mol, and $\theta_{\rm CW} \simeq -21.2$~K) for Sr$_3$CoNb$_2$O$_9$ and Sr$_3$CoTa$_2$O$_9$, respectively.
These values of $C$ correspond to an effective moment ($\mu_{\rm eff} = \sqrt{3k_{\rm B}C/N_{\rm A}}$ where, $N_{\rm A}$ is the Avogadro number and $k_{\rm B}$ is the Boltzmann constant) of $\mu_{\rm eff} \simeq 5.2~\mu_{\rm B}$ and $\sim 5.1~\mu_{\rm B}$, respectively which are close to the expected spin-only value for a $S=3/2$ Co$^{2+}$ ion.

Similarly, the CW fit [Eq.~\eqref{CW}] to the $1/\chi$ data was performed in the low temperature region by varying the fitting range between $20$~K and $60$~K. The obtained parameters are [$C \simeq 1.85(5)$~cm$^3$K/mol and $\theta_{\rm CW} \simeq -7.5(5)$~K] and [$C \simeq 1.93(3)$~cm$^3$K/mol and $\theta_{\rm CW} \simeq -8(1)$~K], for Sr$_3$CoNb$_2$O$_9$ and Sr$_3$CoTa$_2$O$_9$, respectively. During the fitting procedure, $\chi_0$ was fixed to the Van-Vleck susceptibility ($\chi_{\rm VV}$) obtained from the high-field magnetization data (discussed later). To visualize the low temperature linear behaviour, we have plotted $1/(\chi - \chi_{\rm VV})$ vs $T$ in the right $y$-axis for both the compounds. These values of $C$ provide the effective magnetic moment of $\mu_{\rm eff} \simeq 3.84(1)$~$\mu_{\rm B}$ and $3.92(2)$~$\mu_{\rm B}$ for Nb and Ta compounds, respectively which are indeed close to the value expected for $j_{\rm eff}=1/2$, assuming $g=4$. Such a large value of $g$ is not unusual and is typically observed for Co$^{2+}$ systems from the electron-spin-resonance (ESR) experiments due to dominant spin-orbit coupling, at low temperatures~\cite{Susuki267201,Wellm100420}. The negative value of $\theta_{\rm CW}$ indicates that the dominant interaction between the spins is AFM in nature.

In a spin system, $\theta_{\rm CW}$ is a measure of the exchange coupling and is given by $\theta_{\rm CW}=[-zJS(S+1)]/3k_{\rm B}$, where $J$ is the nearest-neighbour(NN) exchange coupling with the Heisenberg Hamiltonian $H = J \sum S_i \cdot S_j$ and $z$ is the number of NN spins~\cite{Domb296}. As both the compounds are having the triangular geometry, we have $z = 6$. Thus, using the value of $\theta_{\rm CW}$, $z$, and $j_{\rm eff}$ in place of $S$ in the above expression, we obtained $J/k_{\rm B} \simeq 5$~K and $5.3$~K for Sr$_3$CoNb$_2$O$_9$ and Sr$_3$CoTa$_2$O$_9$, respectively.

\begin{figure}
	\centering
	\includegraphics[width= \linewidth] {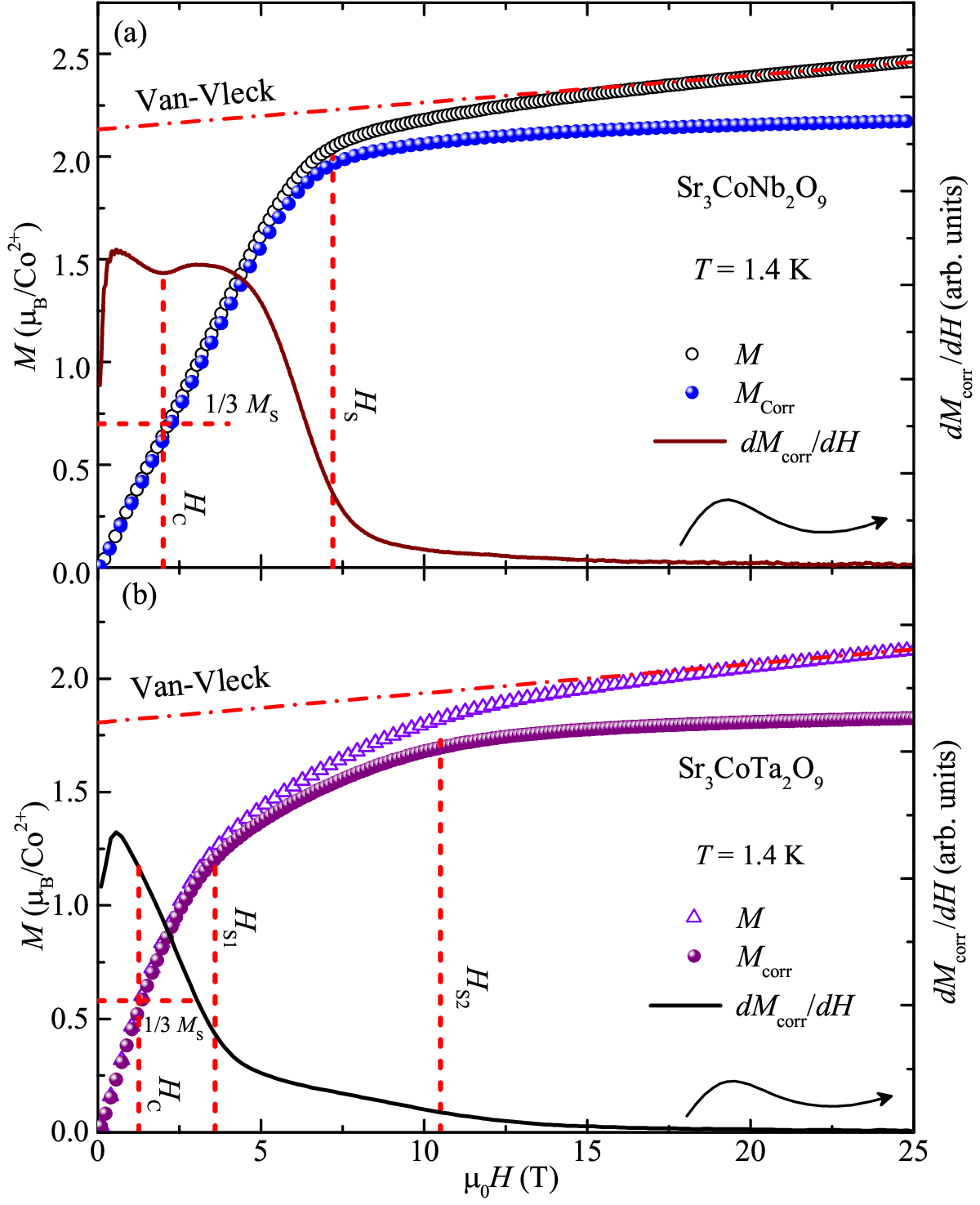}
	\caption{Magnetization ($M$) vs $H$ measured at $T=1.4$~K using pulsed magnetic field for (a) Sr$_3$CoNb$_2$O$_9$ and (b) Sr$_3$CoTa$_2$O$_9$. The dash-dotted line represents the linear fit to the data in high fields which is extrapolated down to zero field to obtain the Van-Vleck magnetization. The corrected magnetization $M_{\rm corr}$ after the subtraction of Van-Vleck contribution is also plotted vs $H$. The dashed lines mark the critical fields. $dM_{\rm corr}/dH$ vs $H$ is plotted in the right $y$-axis to highlight the features at the critical fields.}
	\label{Fig4}
\end{figure}
Further validation of the $j_{\rm eff} =1/2$ ground state and the estimation of exchange coupling were obtained from the high-field magnetization data. The high-field magnetization measured as a function of field upto 60~T using the pulsed magnetic field at the base temperature of $T = 1.4$~K is shown in Fig.~\ref{Fig4}(a) and (b) for Sr$_3$CoNb$_2$O$_9$ and Sr$_3$CoTa$_2$O$_9$, respectively. The pulsed field data are scaled with respect to the magnetization data measured in the SQUID magnetometer up to 7~T at $T=1.8$~K~\footnote{Note that, as the pulse-field magnetization measurements are performed adiabatically at $T=1.4$~K, there would be an adiabatic temperature change ($\Delta T_{\rm ad}$) with increasing field due to magnetocaloric effect (MCE). At $T = 1.4$~K, we calculated $\Delta T_{\rm ad} \sim 0.6$~K which sets an error bar of $T= (1.4 \pm 0.6)$~K. Thus, the scaling of pulse-field data at $T=1.4$~K with respect to the SQUID data at $T=1.8$~K is resonable.}.
For Sr$_3$CoNb$_2$O$_9$, $M$ increases linearly with $H$ and then show a sharp bend at the saturation field $H_{\rm S} \simeq 7.2$~T. Such a sharp bend in the powder sample demonstrates isotropic $g$-factor and/or exchange interaction, as in the case of Ba$_3$Co(Nb,Sb)$_2$O$_9$~\cite{Yokota014403,Susuki267201}. For $H > 8$~T, though it shows the tendency of saturation, still there is a slow increase. This slow increase of magnetization above $H_{\rm S}$ is attributed to the temperature independent Van-Vleck paramagnetism associated with the Co$^{2+}$ ion in the non-cubic environment~\cite{Shirata057205}. The slope of a linear fit for $H > 15$~T and its intercept in the $y$-axis result in the Van-Vleck paramagnetic susceptibility $\chi_{\rm VV} \simeq 7.32\times 10^{-3}$~cm$^{3}$/mol and saturation magnetization $M_{\rm S} \simeq 2.13~\mu_{\rm B}$/Co$^{2+}$, respectively. Using this value of $M_{\rm S}$, the $g$-factor ($M_{\rm S} = gS\mu_{\rm B}$) is calculated to be $g \simeq 4.26$, assuming $j_{\rm eff} = 1/2$. Similar $g$-value is also reported for other Co$^{2+}$ based $j_{\rm eff} = 1/2$ TLAFs~\cite{Shirata057205,Lee104420,Ranjith115804,Rawl060412}. The corrected magnetization ($M_{\rm corr}$) after subtracting the Van-Vleck contribution is also plotted in the same graph. To precisely pinpoint the saturation field we have plotted the derivative $dM_{\rm corr}/dH$ vs $H$ in the right $y$-axis. The curve exhibits a valley at $H \simeq 2$~T which corresponds to $1/3$ of $M_{\rm S}$ and then a sharp drop at $\mu_{\rm 0}H_{\rm S} \simeq 7.2$~T (or, a slope change in the $M_{\rm corr}$ vs $H$ curve) indicating the saturation field or critical field above which the spin system attains the fully polarized state~\cite{Shirata057205}.

Unlike Sr$_3$CoNb$_2$O$_9$, the magnetic behaviour of Sr$_3$CoTa$_2$O$_9$ is found to be somewhat different. Magnetization shows a broad bend between two critical fields $\mu_{\rm 0} H_{\rm S1} \simeq 3.6$~T and $\mu_{\rm 0}H_{\rm S2} \simeq 10.5$~T. Above $H_{\rm S2}$, the magnetization saturates but yet there is a slow increase, similar to Sr$_3$CoNb$_2$O$_9$. A straight line fit above 20~T yields $\chi_{\rm VV} \simeq 7.2\times 10^{-3}$~cm$^{3}$/mol and $M_{\rm S} \simeq 1.8~\mu_{\rm B}$/Co$^{2+}$. This value of $M_{\rm S}$ corresponds to $g \simeq 3.6$ with $j_{\rm eff} = 1/2$. The Van-Vleck corrected magnetization $M_{\rm corr}$ and its derivative $dM_{\rm corr}/dH$ as a function of $H$ are plotted in the left and right $y$-axes, respectively in Fig.~\ref{Fig4}(b) which show pronounced features at $H_{\rm S1}$ and $H_{\rm S2}$. These features are quite different from the broad contort expected due to $g$-factor anisotropy. This is an indication of the existence of an exotic field-induced state between $H_{\rm S1}$ and $H_{\rm S2}$, similar to Na$_2$Co$_2$TeO$_6$~\cite{Lin5559}. No obvious feature is seen at the field corresponding to the $1/3$ magnetization.

The saturation field defines the energy required to overcome the antiferromagnetic exchange energy and polarize the spins in the direction of magnetic field. In particular, in a Heisenberg TLAF, $H_{\rm S}$ can be written in terms of the intralayer exchange coupling as $\mu_{\rm 0}H_{s} = 9JS/g\mu_{\rm B}$~\cite{Kawamura4530}. For Sr$_3$CoNb$_2$O$_9$, our experimental value of $\mu_{\rm 0}H_{\rm s} \simeq 7.2$~T yields an average exchange coupling of $J/k_{\rm B} \simeq 4.7$~K. Similarly, for Sr$_3$CoTa$_2$O$_9$, $\mu_{\rm 0} H_{\rm S2} \simeq 10.5$~T gives $J/k_{\rm B} \simeq 5.8$~K. These values of $J/k_{\rm B}$ are in reasonable agreement with the ones obtained from the analysis of $\theta_{\rm CW}$. The slight difference in magnitude can be attributed to the magnetic anisotropy present in the compounds.


Heisenberg TLAFs typically show a $1/3$ magnetization plateau in an intermediate field range where the spins evolve from a conventional $120^{\degree}$ spin structure to a $uud$ state~\cite{Gvozdikova164209,Seabra214418,Yamamoto057204}. Though a weak anomaly is visible for Sr$_3$CoNb$_2$O$_9$ but it is completely smeared for Sr$_3$CoTa$_2$O$_9$. The absence of $1/3$ magnetization plateau is likely due to the polycrystalline nature of the sample where random orientation of the crystallites average out this effect. Moreover, our $M(H)$ is measured at $T = 1.4$~K (above $T_{\rm N1}$) at which the impact of magnetic anisotropy is minimal. Therefore, $M(H)$ measurement below $T_{\rm N2}$ would reveal these features more clearly.

\subsection{Heat Capacity}
\begin{figure*}
	\centering
	\includegraphics[width= \linewidth] {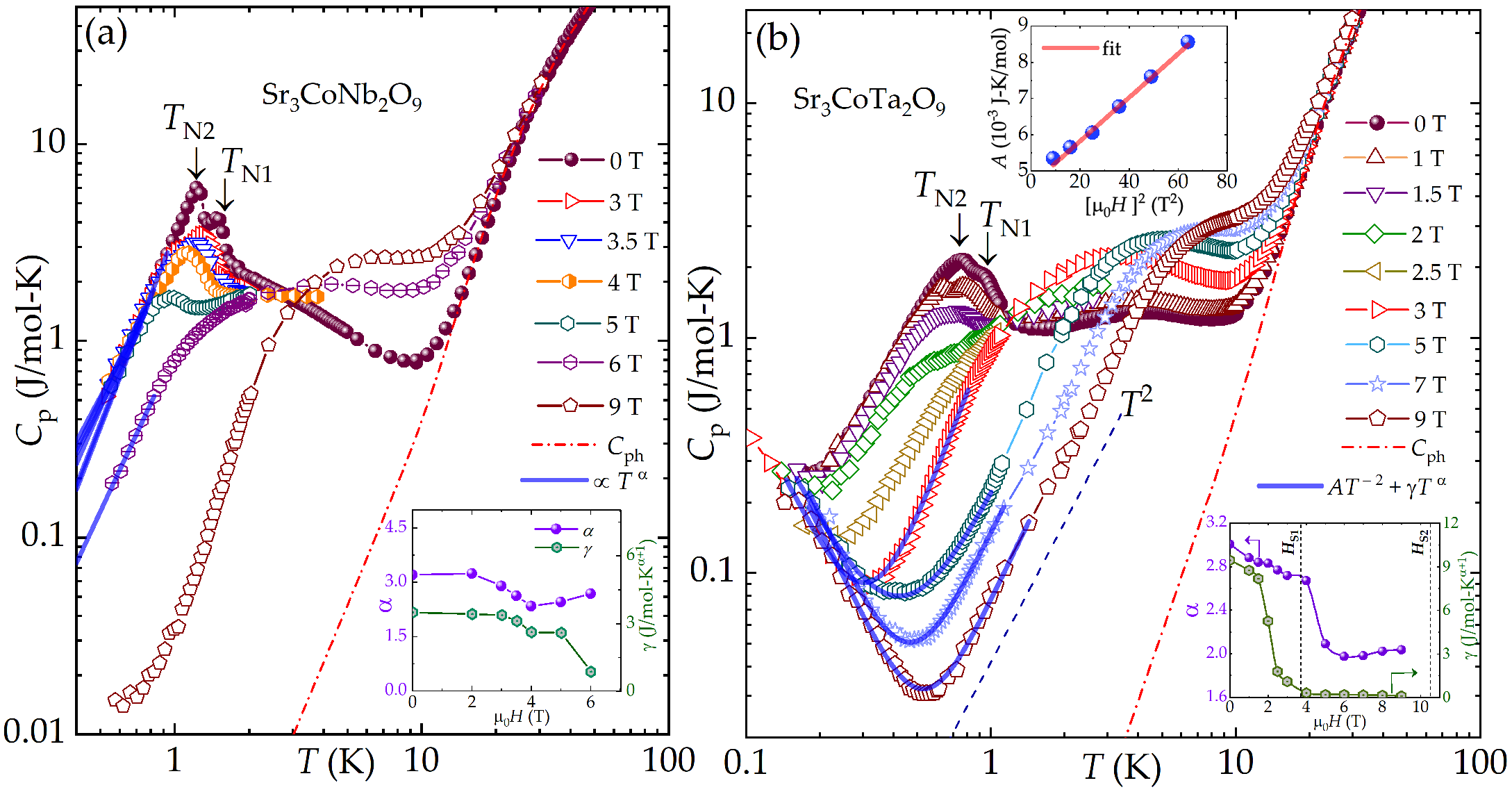}
	\caption{Temperature dependent heat capacity [$C_{\rm p}(T)$] measured in different magnetic fields for (a) Sr$_3$CoNb$_2$O$_9$ and (b) Sr$_3$CoTa$_2$O$_9$. The red dash-dotted line represents the lattice heat capacity using Eq.~\eqref{Eq.2}. The black solid lines are the power law and power law + nuclear Schottky fits to the low-$T$ data of Sr$_3$CoNb$_2$O$_9$ and Sr$_3$CoTa$_2$O$_9$, respectively. The dashed line in (b) marks the power law with $\alpha = 2$. Lower insets: Variation of $\alpha$ and $\gamma$ with $H$ in the left and right $y$-axes, respectively, obtained from the low temperature data fit. Upper inset in (b) is the plot of $A$ vs $H^2$ with solid line being a linear fit.}
	\label{Fig5}
\end{figure*}
\begin{figure*}
	\centering
	\includegraphics[width= \linewidth] {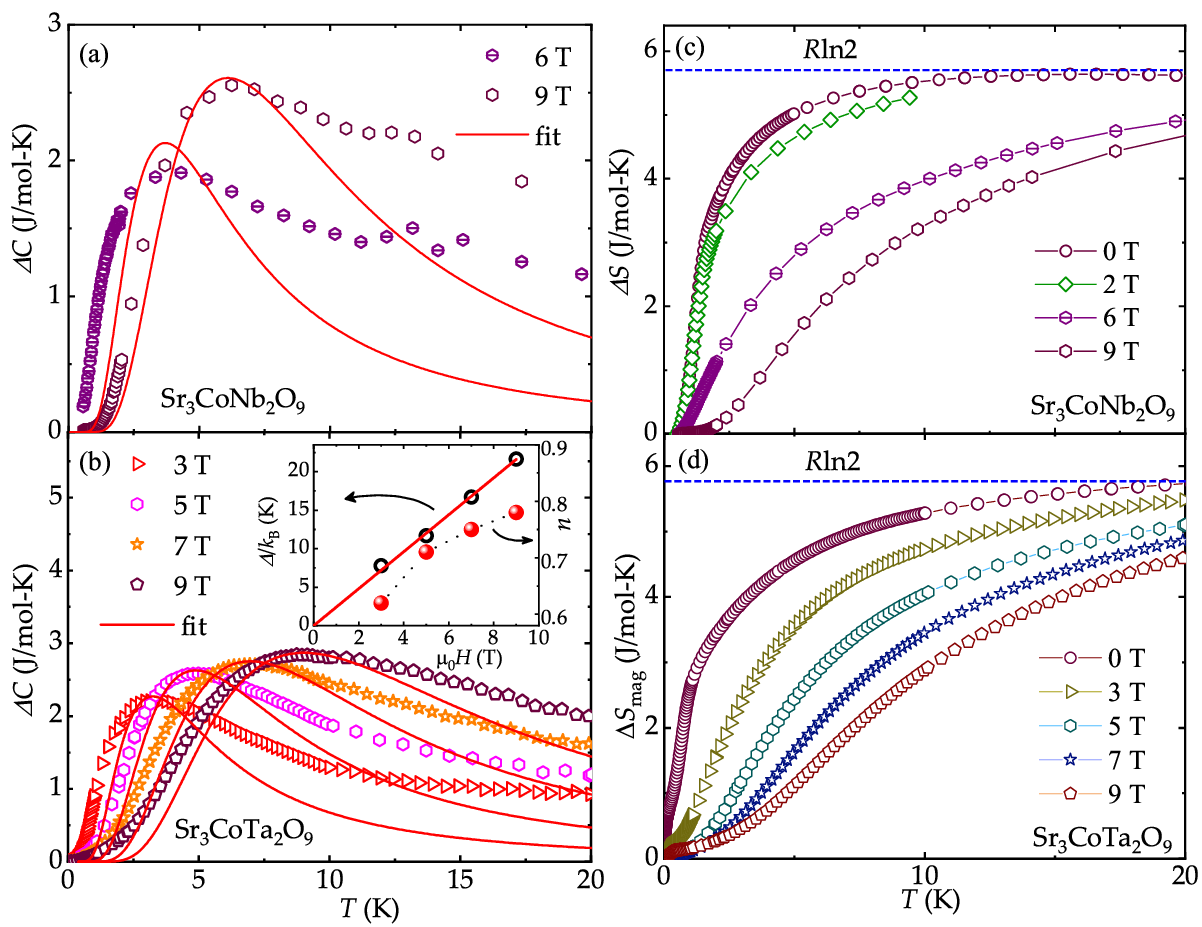}
	\caption{Difference in heat capacity $\Delta C$ ($=C_{\rm p}-C_{\rm ph}-C_{\rm n}$) vs $T$ in different magnetic fields for (a) Sr$_3$CoNb$_2$O$_9$ and (b) Sr$_3$CoTa$_2$O$_9$. The solid lines are the fits using Eq.~\eqref{Eq.5}. Inset of (b) shows the plot of $\Delta/k_{\rm B}$ and $n$ as a function of $H$ for Sr$_3$CoTa$_2$O$_9$ in the left and right $y$-axes, respectively. The change in magnetic entropy $\Delta S$ vs $T$ in different magnetic fields for (c) Sr$_3$CoNb$_2$O$_9$ and (d) Sr$_3$CoTa$_2$O$_9$.}
	\label{Fig6}
\end{figure*}
To delineate the low-energy excitations, temperature-dependent heat capacity [$C_{\rm p}(T)$] measured in different applied fields is shown in Fig.~\ref{Fig5}(a) and (b) for Sr$_3$CoNb$_2$O$_9$ and Sr$_3$CoTa$_2$O$_9$, respectively. The overall temperature variation of $C_{\rm p}$ for both the compounds is found to be nearly same. In zero-field, as the temperature decreases, $C_{\rm p}$ decreases monotonically and below $\sim 10$~K it displays a weak and broad maximum, a possible indication of short-range ordering due to two-dimensionality of the spin-lattice. With further decrease in temperature, two well defined peaks appear at $T_{\rm N1} \simeq 1.47$~K and $T_{\rm N2} \simeq 1.22$~K for Sr$_3$CoNb$_2$O$_9$ and at $T_{\rm N1} \simeq 0.88$~K and $T_{\rm N2} \simeq 0.67$~K for Sr$_3$CoTa$_2$O$_9$, indicating the onset of two successive magnetic transitions at low temperatures. When external magnetic field is applied the height of the peaks is reduced substantially and the peak positions shift towards low temperatures. At higher fields, both the transition peaks are merged into a broad peak and gradually vanishes from the measurement window. This reflects AFM nature of both the transitions. In addition, as the field increases, the broad maxima initially shifts towards low temperatures as expected for a short-range magnetic order. For higher fields, the position of the maxima shifts in the reverse direction (to high temperatures), its height increases, and shows a drastic broadening, reminiscent of a Schottky anomaly due to CEF splitting. Furthermore, the zero-field $C_{\rm p}(T)$ of Sr$_3$CoTa$_2$O$_9$ shows an upturn below $\sim 0.2$~K which moves towards high temperatures with increasing field. This is a typical behaviour of nuclear Schottky arising due to a quadrupole splitting of $^{59}$Co ($I = 7/2$) nuclear levels~\cite{Grivei1301,*An037002}.

In a magnetic insulator, $C_{\rm p}$ in zero-field has three major contributions: lattice heat capacity ($C_{\rm ph}$) due to phonon vibrations which usually dominates at high temperatures, magnetic heat capacity ($C_{\rm mag}$) due to spins which becomes predominant at low temperatures, and the nuclear Schottky contribution ($C_{\rm n}$) which is effective only at very low temperatures (below $\sim 10^{-2}$~K).
In the absence of a non-magnetic analogue, $C_{\rm ph}(T)$ is evaluated by fitting $C_{\rm p}(T)$ in the high temperature region by a linear combination of one Debye and four Einstein terms~\cite{Sebastian064413}
\begin{equation}
	C_{\rm ph}(T) = f_{\rm D}C_{\rm D}(\theta_{\rm D},T)+\sum\limits_{i=1}^{4} g_{i}C_{\rm E_i}(\theta_{{\rm E}_i},T).
	\label{Eq.2}
\end{equation}
The first term in Eq.~\eqref{Eq.2} is the Debye model
\begin{equation}
	C_{\rm D} (\theta_{\rm D}, T)=9nR\left(\frac{T}{\theta_{\rm D}}\right)^{3} \int_0^{\frac{\theta_{\rm D}}{T}}\frac{x^4e^x}{(e^x-1)^2} dx,
	\label{Eq.3}
\end{equation}
where $R$ is the universal gas constant, $n$ is the number of atoms in the formula unit, and $\theta_{\rm D}$ is the characteristic Debye temperature. The flat optical modes in the phonon spectra are accounted for by the second term in Eq.~\eqref{Eq.2}, called the Einstein term
\begin{equation}
	C_{\rm E}(\theta_{\rm E}, T) = 3nR\left(\frac{\theta_{\rm E}}{T}\right)^2 
	\frac{e^{\left(\frac{\theta_{\rm E}}{T}\right)}}{[e^{\left(\frac{\theta_{\rm E}}{T}\right)}-1]^{2}},
	\label{Eq.4} 
\end{equation}
where $\theta_{\rm E}$ is the characteristic Einstein temperature.
The coefficients $f_{\rm D}$, $g_1$, $g_2$, $g_3$, and $g_4$ are the weight factors, which take into account the number of atoms per formula unit ($n$) and are chosen in such a way that $f_{\rm D} + g_1 + g_2 + g_3 + g_4 = 1$ and the Dulong-Petit value ($\sim 3nR$) is satisfied at high temperatures.

The zero-field $C_{\rm p}(T)$ above 40~K is fitted by Eq.~\eqref{Eq.2} and the obtained fitting parameters are ($f_{\rm D} \simeq 0.066$, $g_1 \simeq 0.066$, $g_2 \simeq 0.20$, $g_3 \simeq 0.13$, $g_4 \simeq 0.533$, $\theta_{\rm D} \simeq 178$~K, $\theta_{\rm E1} \simeq 117$~K, $\theta_{\rm E2} \simeq 206$~K, $\theta_{\rm E3} \simeq 328$~K and $\theta_{\rm E4} \simeq 520$~K) and ($f_{\rm D} \simeq 0.066$, $g_1 \simeq 0.066$, $g_2 \simeq 0.20$, $g_3 \simeq 0.13$, $g_4 \simeq 0.533$, $\theta_{\rm D} \simeq 163$~K, $\theta_{\rm E1} \simeq 117$~K, $\theta_{\rm E2} \simeq 195$~K, $\theta_{\rm E3} \simeq 294$~K and $\theta_{\rm E4} \simeq 490$~K) for Sr$_{3}$CoNb$_{2}$O$_{9}$ and Sr$_{3}$CoTa$_{2}$O$_{9}$, respectively. The fit was extrapolated down to the lowest measured temperature (see Fig.~\ref{Fig5}) to obtain $C_{\rm ph}(T)$. The estimation of $C_{\rm n}$ is discussed later. The magnetic and/or Schottky contributions, $\Delta C$, obtained by subtracting $C_{\rm ph}$ and $C_{\rm n}$ from total $C_{\rm p}$ are plotted as a function of $T$ for different fields in Fig.~\ref{Fig6}(a) and (b) for the Nb and Ta compounds, respectively. The respective change in magnetic entropies evaluated by integrating $\Delta C/T$ with respect to $T$ [i.e. ${\Delta S}=\int\limits_{0}^{T} \frac{\Delta C (T^{\prime})}{T^{\prime}}dT^{\prime}$] for different fields are shown in Fig.~\ref{Fig6}(c) and (d). The saturation value of $\Delta S$ approaches $R\ln2$ above $\sim 30$~K irrespective of the applied field for both the compounds, further evidencing $j_{\rm eff} = 1/2$ ground state for Co$^{2+}$.

$\Delta C(T)$ in different fields can be fitted by the following two-level Schottky function
\begin{equation}
	C_{\rm S} = n_{\rm S}R \left(\frac{\Delta}{k_{B}T}\right)^2 \frac{e^{-\frac{\Delta}{k_{\rm B}T}}} {\left (1+e^{-\frac{\Delta}{k_{\rm B}T}} \right)^2},
	\label{Eq.5}
\end{equation}
where $n_{\rm S}$ is the number of free spins per f.u. contributing to the Schottky behaviour and $\Delta/k_{\rm B}$ is the Zeeman gap in magnetic fields. We fitted the data in different magnetic fields by Eq.~\eqref{Eq.5} making $n$ and $\Delta/k_{\rm B}$ as fitting parameters [Fig.~\ref{Fig6}(a) and (b)]. The fit reproduces the experimental data reasonably well, especially in the broad maximum regime. However, the deviation in the high temperature regime can be ascribed to the unreliable subtraction of phonon contribution and the presence of magnetic contribution. The obtained $\Delta/k_{\rm B}$ and $n$ are plotted as a function of field in the inset of Fig.~\ref{Fig6}(b) for Sr$_3$CoTa$_2$O$_9$. With increasing $H$, $n$ increases systematically suggesting the excitation of more free spins to the higher energy levels. It is expected to reach the maximum value ($\sim 1$) in the fully polarized state, above $H_{\rm S2} \simeq 10.5$~T. Similarly, $\Delta/k_{\rm B}$ varies linearly with $H$.
Taking the value of $\Delta/k_{\rm B} \simeq 21.65$~K at $H = 9$~T which is close to the saturation field, we obtained $g \simeq 3.58$. This value of $g$ is indeed close to the one obtained from the magnetization analysis. This further elucidates that the Schottky effect in heat capacity is arising from the Co$^{2+}$ Kramers' doublets with $j_{\rm eff} = 1/2$.

\section{Discussion}
\begin{figure}
	\centering
	\includegraphics[width= \linewidth] {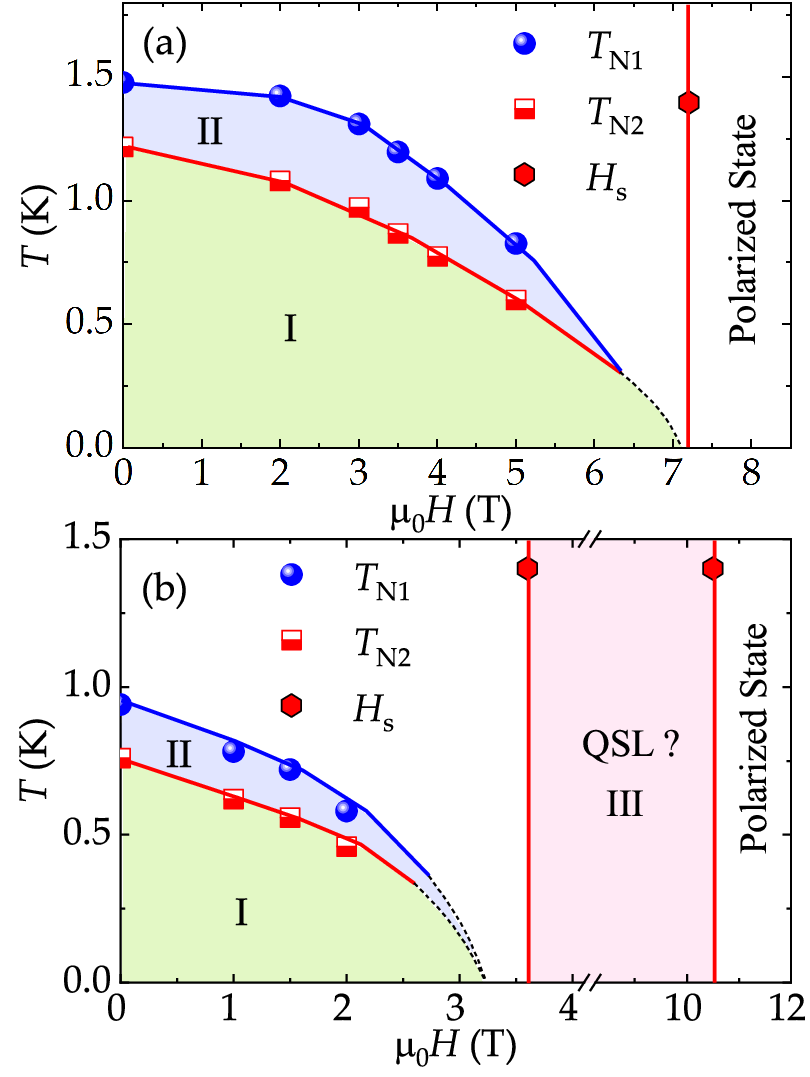}
	\caption{$T$ vs $H$ phase diagram for (a) Sr$_{3}$CoNb$_{2}$O$_{9}$ and (b) Sr$_{3}$CoTa$_{2}$O$_{9}$ constructed using the heat capacity and magnetization data. The exotic field induced regime (III) between $H_{\rm S1}$ and $H_{\rm S2}$ for Sr$_{3}$CoTa$_{2}$O$_{9}$ is highlighted in a different color.}
	\label{Fig7}
\end{figure}
For both the compounds, double magnetic transitions are observed at $T_{\rm N1}$ and $T_{\rm N2}$ ($T_{\rm N2} < T_{\rm N1}$) in zero-field. It is predicted that double transitions can occur in TLAFs when the magnetic anisotropy is of easy-axis type, while the TLAFs with easy-plane type anisotropy may show only single transition~\cite{Matsubara2424,Miyashita3385,Quirion014414,Ranjith014415}. On lowering the temperature, the collinear $uud$ state appears before $120\degree$ state in TLAFs with easy-axis anisotropy. When magnetic field is applied, an additional high-field 2:1 canted phase is also stabilized~\cite{Seabra214418}. The temperature range of the intermediate phase ($T_{\rm N1}-T_{\rm N2})/T_{\rm N1}$ reflects the relative strength of easy-axis anisotropy with respect to the isotropic intralayer coupling.
Similar type of double transitions are reported for Ba$_3$Co(Nb,Ta,Sb)$_2$O$_9$ and some other compounds~\cite{Lee104420,Zhou267206,Ranjith115804,Yokota014403,Zhou267206}.
Thus, the two consecutive magnetic transitions at low temperatures is attributed to the easy-axis anisotropy in both our compounds and the $120\degree$ ordering below $T_{\rm N2}$ is preceded by a collinear order between $T_{\rm N1}$ and $T_{\rm N2}$. The observed narrow temperature regime between $T_{\rm N1}$ and $T_{\rm N2}$ in both the compounds suggests considerably weak anisotropy compared to the isotropic exchange coupling in the Hamiltonian~\cite{Shirata057205}. This implies that these compounds are more close to Heisenberg model in contrast to strong anisotropy anticipated for Co-based compounds~\cite{Line546,*Shiba2326} and the local environment of Co$^{2+}$ is close to a cubic symmetry as in KCoF$_3$ and Ba$_3$CoSb$_2$O$_9$~\cite{Shirata057205}. Recent theoretical studies claim that the Heisenberg model on a triangular lattice with no anisotropy can also show double phase transition~\cite{Seabra214418}. It is also suggested that the transition from $uud$ to $120^0$ state (at $T_{\rm N2}$) can be described by the conventional Berezinskii-Kosterlitz-Thouless (BKT) universality class.
The magnetic ($T-H$) phase diagram constructed using the transition temperatures from $C_{\rm p}(T)$ and saturation fields from the magnetization data are presented in Fig.~\ref{Fig7}(a) and (b) for Sr$_3$CoNb$_2$O$_9$ and Sr$_3$CoTa$_2$O$_9$, respectively.
Both the compounds exhibit two common phase regimes I and II known to be $120\degree$ and collinear spin states, respectively, as typically observed in majority of the TLAFs.
On increasing field, both the transitions are merged into one and are expected to approach zero temperature at around $\sim 7$~T for Sr$_3$CoNb$_2$O$_9$ and $\sim 3$~T for Sr$_3$CoTa$_2$O$_9$, respectively. These fields are very close to their respective saturation fields.

The most striking difference between these two compounds is that Sr$_3$CoTa$_2$O$_9$ shows the appearance of an exotic phase in an extended field range [regime III in Fig.~\ref{Fig7}(b)] while for Sr$_3$CoNb$_2$O$_9$ this phase is either absent or exists over a narrow field range. In order to understand the peculiar behaviour in the magnetization data, we fitted $C_{\rm p}(T)$ in the low temperature regime by a power law of the form $C_{\rm p} = \gamma T^{\alpha}$ for Sr$_3$CoNb$_2$O$_9$. To accommodate the low-$T$ upturn, $C_{\rm p}(T)$ data of Sr$_3$CoTa$_2$O$_9$ were fitted with a sum of $A/T^2$ and $\gamma T^{\alpha}$ from the lowest temperature to 1~K~\cite{Ding144432}.
Here, $A/T^2$ accounts for the high temperature part of the nuclear Schottky anomaly, which is nothing but the high temperature approximation of Eq.~\eqref{Eq.5}~\cite{Gopal2012}. The constant $A$ is related to the nuclear level splitting $\Delta$ (both quadrupolar and Zeeman) with $A \propto \Delta ^2$ for $T >> \Delta$. $A$ is found to increase quadratically with $H$ [upper inset of Fig.~\ref{Fig5}(b)], which is consistent with the theoretical prediction~\cite{An037002}.

The obtained values of $\gamma$ and $\alpha$ are plotted against $H$ in the lower insets of the respective figures. For Sr$_3$CoNb$_2$O$_9$, the value of $\alpha$ is found to be close to 3 for all the measured fields [inset of Fig.~\ref{Fig5}(a)], as expected in a 3D AFM ordered state. Only at $H = 9$~T which is well above $H_{\rm S}$, the fitted parameters were not reliable, possibly because of large scattering in the low temperature data and low heat capacity values.
On the other hand, for Sr$_3$CoTa$_2$O$_9$, the value of $\alpha$ is found to be close to 3 (with $\gamma \simeq 1042$~mJ/mol.K$^{3.75}$) for $H=3$~T, as expected [lower inset of Fig.~\ref{Fig5}(b)]. For $H > 4$~T, it is reduced significantly to about $\sim 2$ and remains almost field independent. Interestingly, these fields fall within the exotic regime (between $\mu_{0}H_{\rm S1} \simeq 3.6$~T and $\mu_{0}H_{\rm S2} \simeq 10.5$~T) found from the magnetization measurements. In these fields, the value of $\gamma$ varies from $\sim 300$ to 90~mJ/mol.K$^{3}$ which is a significant reduction from the 3~T value but it is still considerably large.

Typically, one expects a $T^3$ behaviour for $C_{\rm p}(T)$ due to 3D spin-wave dispersion in the AFM ordered state~\cite{Nath024431}. On the other hand, a more conventional characteristic feature of QSL is the linear temperature dependency of low temperature $C_{\rm p}(T)$ i.e. $C_{\rm p} = \gamma T$ where, $\gamma$ is related to the spinon density of states at the spinon Fermi surface.
Likewise, a quadratic temperature dependency ($C_{\rm p} \propto T^2$) is predicted theoretically for the gapless Dirac QSLs at low temperatures~\cite{Ran117205,Hermele224413}. Indeed, several QSL candidates with triangular, kagome, and hyperkagome geometries are reported to evince such a behaviour at low temperatures~\cite{Kundu267202,nakatsuji1697,Helton107204,Okamoto137207,Ding144432}.
Thus, the suppression of magnetic LRO and power law behaviour of $C_{\rm p}$ with a reduced value of $\alpha \sim 2$ in the critical field region unambiguously point towards an exotic phase regime, possibly the field induced QSL in Sr$_3$CoTa$_2$O$_9$~\cite{Clark207208,cheng197204}. Moreover, the obtained large value of $\gamma$ can also be considered as a generic feature of low-energy gapless excitations, as observed for gapless QSL candidates Ba$_3$CuSb$_2$O$_9$ and Sr$_2$Cu(Te$_{0.5}$W$_{0.5}$)O$_6$~\cite{Zhou147204,*Mustonen1085}.

This type of field induced behaviour is observed in some honeycomb lattices with strong Kitaev interaction. The Kitaev physics demands the presence of bond-dependent anisotropy which can be realized in honeycomb lattices as well as spin dimers with strong spin-orbit coupling where the metal octahedra are either edge shared or parallelly edge shared~\cite{Trebst1,*Jackeli017205}. Further, it is predicted that spin-orbit coupled honeycomb systems with $j_{\rm eff} = 1/2$ are ideal boulevard to host a Kitaev spin liquid~\cite{Takagi280}. Indeed, $j_{\rm eff} = 1/2$ honeycomb lattices $\alpha$-RuCl$_3$~\cite{Baek037201}, Na$_2$Co$_2$TeO$_6$~\cite{Lin5559}, BaCo$_{2}$(AsO$_{4}$)$_{2}$~\cite{Zhong6953}, and BaCo$_{2}$(P$_{\rm 1-x}$V$_{x}$)O$_{8}$~\cite{Zhong220407} with strong anisotropy show the onset of conventional magnetic LRO at low temperatures. Under magnetic field, the LRO is suppressed (i.e. the anisotropic bond dependent coupling dominates over the isotropic coupling) and the system eventually crosses over to a low temperature non-magnetic or disordered state which is understood to be the field induced Kitaev spin-liquid~\cite{Trebst1}.
A triangular lattice can also be viewed as a superposition of honeycomb layers in certain stacking sequence. This analogy can be extended to the TLAFs Sr$_3$Co(Nb,Ta)$_2$O$_9$. A careful inspection of the crystal structure reveals
that the CoO$_6$ octahedra in both the compounds are parallel edge shared, when the non-magnetic (Nb,Ta)O$_6$ octahedra are removed.
In Fig.~\ref{Fig1}(b), we have schematized the ways in which parallel edge sharing could be feasible in Sr$_3$Co(Nb,Ta)$_2$O$_9$. As one can clearly see that the identical Co-octahedra are perfectly parallel edge shared, which might facilitate a bond-dependent anisotropy and hence Kitaev interaction~\cite{Wellm100420,Maksimov021017}. Meanwhile, due to low symmetry crystal structure, there is an inherent distortion in the metal octahedra due to which the edge sharing between two dissimilar octahedra [Co(1)O$_6$ and Co(2)O$_6$] is deviating slightly from the perfect parallel edge sharing. It is to be noted that the recent theoretical work on TLAFs has also predicted a transition from an ordered magnetic state in the isotropic case into a QSL state in the easy-axis regime, with increasing easy-axis anisotropy~\cite{Ulaga2023}. Nevertheless, to conclusively establish this novel phenomena in these TLAFs, inelastic neutron scattering and $\mu$SR experiments in dilution temperatures and under magnetic fields on good quality single crystals would be imperative.

Experimentally, the extent of frustration in any magnetically frustrated material is quantified by the frustration ratio $f= {|\theta_{\rm CW}|}/{T_{\rm N}}$~\cite{Ramirez453}. From the value of $\theta_{\rm CW}$ and $T_{\rm N1}$, the frustration parameter is calculated to be $f \sim 5$ and $\sim 9$ for Sr$_{3}$CoNb$_{2}$O$_{9}$ and Sr$_{3}$CoTa$_{2}$O$_{9}$, respectively. Despite same crystal structure, the latter shows stronger frustration and a fluctuating regime over an extended field range while for the former, the extent of frustration is less and hence the fluctuating regime is narrowed.
The possible origin of such a drastic difference in magnetic properties can be attributed to the hierarchy of the exchange couplings. It is to be noted that the Co$^{2+}$~-~Co$^{2+}$ superexchange involves Co$^{2+}(3d)$~-~O$^{2-}(2p)$~-~Nb$^{5+}(4p)$~-~O$^{2-}(2p)$~-~Co$^{2+}(3d)$ and Co$^{2+}(3d)$~-~O$^{2-}(2p)$~-~Ta$^{5+}(4f)$~-~O$^{2-}(2p)$~-~Co$^{2+}(3d)$ pathways for Sr$_3$CoNb$_2$O$_9$ and Sr$_3$CoTa$_2$O$_9$, respectively. Hence, the difference is coupling strength is likely due to the participation of different orbitals of Nb$^{5+}(4p)$ and Ta$^{5+}(4f)$ in the interaction mechanism~\cite{Yokota014403,Watanabe054414}.

\section{Conclusion}
We report a detailed study of the thermodynamic properties of two new quantum magnets Sr$_3$Co(Nb,Ta)$_2$O$_9$ possessing Co$^{2+}$ triangular layers. Their low temperature properties are unequivocally described by $j_{\rm eff}=1/2$ Kramers doublet of Co$^{2+}$ ions interacting antiferromagnetically.
Similar to many other TLAFs with easy-axis anisotropy, both the compounds undergo two sequential magnetic transitions at low temperatures. Sr$_3$CoNb$_2$O$_9$ exhibits a weak plateau type feature at the $1/3$ magnetization while this feature is smeared for Sr$_3$CoTa$_2$O$_9$, likely due to the effect of magnetic anisotropy and/or random orientation of the crystallites. Despite same crystal structure, Sr$_3$CoTa$_2$O$_9$ with $f \simeq 9$ is found to be more frustrated than Sr$_3$CoNb$_2$O$_9$ with $f \simeq 5$, which can be attributed to the involvement of different orbitals of Ta$^{5+}$ and Nb$^{5+}$ in the interaction paths. The existence of an exotic field induced quantum phase is detected for Sr$_3$CoTa$_2$O$_9$ over a wide field range, below the fully polarized state while this behaviour is either absent or exists over a narrow field range for the Nb analogue. Because of this non-trivial character, Sr$_3$CoTa$_2$O$_9$ is a model system for further experimental as well as theoretical investigations at low temperatures and under magnetic fields.

$Note~added$: During the course of our work, we became aware of an independent study on Sr$_3$CoTa$_2$O$_9$ [crystal structure: trigonal ($P \bar{3}m1$)]~\cite{Nishizawa174419}.
They have also observed double transitions at low temperatures.

\section*{Acknowledgements}
We would like to acknowledge SERB, India, for financial support bearing sanction Grant No.~CRG/2022/000997. We also acknowledge the support of HLD-HZDR, member of the European Magnetic Field Laboratory (EMFL). S.J.S. is supported by the Prime Minister’s Research Fellowship (PMRF) scheme, Government of India.

%

\end{document}